\documentclass[aps,floatfix,amsmath,nofootinbib,amssymb,superscriptaddress]{revtex4}

\usepackage{overpic}
\usepackage{amssymb}
\usepackage{indentfirst}
\usepackage{feynmf}   
\usepackage{slashed}  
\usepackage{cases}
\usepackage{color}
\usepackage{multirow}
\usepackage{epstopdf}
\usepackage{graphicx,color,bm}
\usepackage{epstopdf}

\usepackage[colorlinks,
            citecolor=blue,
            anchorcolor=red,
            menucolor=red,
            linkcolor=red,
            filecolor=red,
            runcolor=red,
            urlcolor=blue,
            frenchlinks=red]{hyperref}

\begin{document}

\title{The $\sin (2\phi-\phi_S)$ azimuthal asymmetry in the pion induced Drell-Yan process within TMD factorization}
\author{Hui Li}
\affiliation{School of Physics, Southeast University, Nanjing 211189, China}
\author{Xiaoyu Wang}
\email{xiaoyuwang@zzu.edu.cn}
\affiliation{School of Physics and Engineering, Zhengzhou University, Zhengzhou, Henan 450001, China}
\author{Zhun Lu}
\email{zhunlu@seu.edu.cn}
\affiliation{School of Physics, Southeast University, Nanjing 211189, China}

\begin{abstract}

We investigate the single transverse-spin asymmetry with a $\sin (2\phi-\phi_S)$ modulation in the pion-induced Drell-Yan process within the theoretical framework of the transverse momentum dependent factorization.
The asymmetry is contributed by the convolution of the Boer-Mulders function and the transversity.
We adopt the model results for the distributions of the pion meson from the light-cone wavefunction approach and the available parametrization for the distributions of the proton to numerically estimate the $\sin (2\phi-\phi_S)$ asymmetry in $\pi^- p$ Drell-Yan at the kinematics of the COMPASS at CERN.
To implement the TMD evolution formalism of parton distribution functions, we apply the recently extracted nonperturbative Sudakov form factor associated with the distribution functions of the proton and the pion. It is found that our prediction on the single transverse-spin dependent asymmetry ${\sin (2\phi-\phi_S)}$ as functions of $x_p$, $x_\pi$, $x_F$ and $q_\perp$ is qualitatively consistent with the recent COMPASS measurement in both sign and magnitude.
\end{abstract}

\maketitle

\section{Introduction}

The Boer-Mulders function, denoted by $h_1^\perp$, is one of the eight transverse momentum dependent (TMD) parton distribution functions (PDFs) describing the partonic structure of hadrons at leading-twist level.
It represents the transversely polarization asymmetry of quarks inside an
unpolarized hadron~\cite{Boer:1997nt,Boer:1999mm} arising from the correlation between the quark spin and the quark transverse momentum, thereby it manifests the novel structure of hadrons.
However, the very existence of the Boer-Mulders function was not so obvious. Similar to its chiral-even partner--the Sivers function $f_{1T}^\perp$~\cite{Sivers:1989cc}, the Boer-Mulder function was initially thought to vanish under the constraint of (naive) time reversal invariance of QCD~\cite{Collins:1992kk}.
The situation was changed after explicit model calculations~\cite{Brodsky:2002cx,Brodsky:2002rv,Boer:2002ju} incorporating gluon exchange between the struck quark and the spectator shows that the T-odd distributions can actually survive.
The crucial ingredient in the argument is the Wilson lines (or the gauge links) appearing in the full gange-invariant definition of TMDs~\cite{Collins:2002kn,Ji:2002aa}.
The presence of the Wilson lines also indicates that the T-odd distributions, such as Sivers function and the Boer-Mulders function are process dependent, i.e., they change sign~\cite{Collins:2002kn,Brodsky:2002rv,Boer:2002ju} between the semi-inclusive deeply inelastic scattering and Drell-Yan process, a vital prediction which needs verification by future experimental measurement.
In the last decades, the Boer-Mulders function of the proton as well as that of the pion has been studied intensively in literature~\cite{Boer:2002ju,Gamberg:2003ey,Yuan:2003wk,Pobylitsa:2003ty,Bacchetta:2003rz,
Lu:2004au,Lu:2006ew,Gamberg:2007wm,Burkardt:2007xm,Bacchetta:2008af,
Zhang:2008nu,Meissner:2008ay,Courtoy:2009pc,Gamberg:2009uk,Lu:2009ip,Barone:2009hw,Barone:2010gk,
Pasquini:2010af,Lu:2012hh,Pasquini:2014ppa,Lu:2016pdp,Wang:2017onm} by models and phenomenological analysis.

As the Boer-Mulders function is a chiral-odd distribution function, it has to be coupled with another chirlal-odd distribution/fragmentation function to survive in a high energy scattering process.
A promising process for accessing the Boer-Mulders function is the unpolarized Drell-Yan process, which displays an azimuthal dependence of the final-state dilepton with the $\cos 2\phi$ modulation. As proposed by Boer~\cite{Boer:1999mm}, the coupling of two Boer-Mulder function from each incident hadrons can generate such asymmetry. However, recently studies based on higher order perturbative QCD~\cite{Peng:2015spa,Lambertsen:2016wgj,Chang:2018pvk} show that gluon radiation in hard scattering can also give rise to the $\cos2\phi$ asymmetry substantially, making the extraction of the Boer-Mulders function rather difficult.
In the unpolarized semi-inclusive deep inelastic scattering, the combination of the Boer-Mulders function and the Collins fragmentation function $H_1^\perp$ can lead to a similar $\cos 2\phi_h$ azimuthal asymmetry of the final state spin-0 hadron. But this asymmetry is contaminated by the so-call Cahn effect\cite{cahn,Barone:2005kt,Barone:2008tn}, which is a higher-twist kinematical effect due to the transverse motion of the unpolarized quarks.
A cleaner process for accessing the Boer-Mulders function is the single transversely polarized Drell-Yan.
In this process, the convolution of the Boer-Mulders function and the transversity distribution $h_1$ can give rise to a $\sin (2\phi-\phi_S)$ asymmetry~\cite{Boer:1999mm,Arnold:2008kf} with $\phi_S$ the azimuthal angle of target transverse spin.
This makes the transversity function an ideal probe in analysing the information of the Boer-Mulders function from single transversely polarized Drell-Yan because of less contribution from the background.
Recently, the first measurement on the $\sin (2\phi-\phi_S)$ asymmetry has been performed by the COMPASS~\cite{Aghasyan:2017jop}, which adopted a pion beam to collide on the transversely polarized nucleon target.
Although no clear tendency is observed on the $\sin (2\phi-\phi_S)$ asymmetry due to relatively large statistical uncertainties, it indeed indicates negative sign and substantial size.

In this work, we will study the estimate the $\sin (2\phi-\phi_S)$ asymmetry of the pion-nucleon Drell-Yan process by considering the convolution $h_1^\perp \otimes h_1$.
The main purpose is to investigate the feasibility of accessing the Boer-Mulders function from single polarized Drell-Yan.
The theoretical tool we adopt in this study is the TMD factorization~\cite{Collins:1981uk,Collins:1984kg,Collins:2011zzd,Ji:2004xq} which is applicable in the region the transverse momentum of the dilepton $q_\perp$ is much smaller than the hard scale $Q$.
The TMD factorization has been widely applied to various high energy processes, such as the semi-inclusive deep inelastic scatering (SIDIS)~\cite{Collins:1981uk,Collins:2011zzd,Ji:2004wu,Aybat:2011zv,Collins:2012uy,Echevarria:2012pw},
$e^+ e^-$ annihilation~\cite{Collins:2011zzd,Pitonyak:2013dsu,Boer:2008fr}, Drell-Yan~\cite{Collins:2011zzd,Arnold:2008kf} and W/Z production in hadron collision~\cite{Collins:2011zzd,Collins:1984kg,Lambertsen:2016wgj}.
The TMD factorization can be also extended to the moderate $q_\perp$ region where an equivalence~\cite{Ji:2006ub,Ji:2006vf} between the TMD factorization and the twist-3 collinear factorization is found.
From the perspective of TMD factorization, the physical observables in the region $q_\perp \ll Q$ can be expressed as the convolution of the factors related to hard scattering and the well-defined TMD distributions or fragmentation functions~(collectively called as TMDs).
One of the main features of the TMD formalism is that it provides a systematic approach to deal with the evolution of the TMDs.
In this formalism, the energy evolution (or the scale dependence) of TMDs are governed by the so-called Collins-Soper equation~\cite{Collins:1981uk,Collins:1984kg,Collins:2011zzd,Idilbi:2004vb}.
The solution of the evolution equation shows the changes of TMDs from a initial scale to another scale may be determined by an exponential form of the Sudakov-like form factor~\cite{Collins:1984kg,Collins:2011zzd,Aybat:2011zv,Collins:1999dz}, which can be separated to the perturbative part and nonperturbative part.
The former one is perturbatively calculable, while the later one is usually by phenomenological extraction from experimental data.
In this paper, we wil consider the evolution of both the the pion Boer-Mulders function and the proton transversity to estimate the $\sin (2\phi-\phi_S)$ asymmetry at the kinematics of COMPASS and compare the results with recent COMPASS measurement.

The rest of the paper is organized as follows. In Sec.~\ref{Sec.evolution}, we provide a detailed review on the TMD evolution formalism for the unpolarized and polarized TMDs involved in the calculation. particularly, we will present our choice on the nonperturbative Sudakov form factors associated with the TMDs.
In Sec.~\ref{Sec.formalism}, we derive the theoretical expression of ${\sin (2\phi-\phi_S)}$ asymmetry in the pion-nucleon Drell-Yan within the framework of TMD factorization.
In Sec.~\ref{Sec.numerical}, we estimate the asymmetry at the COMPASS kinematics using a model result of the pion Boer-Mulders function and a parametrization for proton transversity as input. We also provide some discussion based on our numerical result.
We summarize the paper in Sec.~\ref{Sec.conclusion}.

\section{The TMD evolution of distribution functions}
\label{Sec.evolution}

In this section, we review the evolution formalism of the unpolarized distribution function $f_1$, the Boer-Mulders function $h_1^{\perp}$ of the pion as well as the transversity function $h_1$ of the proton, within the TMD factorization.

TMD evolution is usually performed in the coordinate $b_\perp$-space, where $b_\perp$ is conjugated to $k_\perp$ in the transverse momentum space via Fourier transformation~\cite{Collins:1984kg,Collins:2011zzd}. One of the main advantages of $b_\perp$-space is that the cross section can be expressed as the production of $b_\perp$-dependent functions instead of the complicate convolution of functions in $k_\perp$-space.
In the TMD factorization based on different schemes (such as the CS-81~\cite{Collins:1981uk}, JMY~\cite{Ji:2004xq,Ji:2004wu} and Collins-11 schemes~\cite{Collins:2011zzd}), the TMD distribution functions $\tilde{F}(x,b;\mu,\zeta_F)$ in $b_\perp$ space depend on two energy scales. One is the renormalization scale $\mu$ which is related to the corresponding collinear PDFs, the other is the energy scale $\zeta_F$ serving as a cutoff to regularize the light-cone singularity in the operator definition of the TMD distributions. The two energy dependencies are encoded in different evolution equations. For the $\zeta_F$ dependence of the TMD distributions, it is determined by the Collins-Soper~(CS) equation ~\cite{Collins:1981uk} ($b= |\bm b_\perp|$):
\begin{align}
\frac{\partial\ \mathrm{ln} \tilde{F}(x,b;\mu,\zeta_F)}{\partial\ \sqrt{\zeta_F}}=\tilde{K}(b;\mu),
\end{align}
while the $\mu$ dependence is derived from the renormalization group equation as
\begin{align}
&\frac{d\ \tilde{K}}{d\ \mathrm{ln}\mu}=-\gamma_K(\alpha_s(\mu)),\\
&\frac{d\ \mathrm{ln} \tilde{F}(x,b;\mu,\zeta_F)}
{d\ \mathrm{ln}\mu}=\gamma_F(\alpha_s(\mu);{\frac{\zeta^2_F}{\mu^2}}),
\end{align}
with $\tilde{K}$ the CS evolution kernel, and $\gamma_K$ and $\gamma_F$ the anomalous dimensions.
Solving those equations, one can obtain the general solution for the energy dependence of $\tilde{F}$:
\begin{equation}
\tilde{F}(x,b,Q)=\mathcal{F}\times e^{-S(Q,b)}\times \tilde{F}(x,b,\mu_i),
\label{eq:fs}
\end{equation}
where $\mathcal{F}$ is the factor related to the hard scattering, $S(Q,b)$ is the Sudakov form factor. Hereafter, we will set $\mu=\sqrt{\zeta_F}=Q$, and express $\tilde{F}(x,b;\mu=Q,\zeta_F=Q^2)$ as $\tilde{F}(x,b;Q)$ for simplicity.
Eq.~(\ref{eq:fs}) demonstrates that the distribution $\tilde{F}$ at an arbitrary scale $Q$ can be determined by the same distribution at an initial scale $\mu_i$ through the evolution encoded by the exponential form $\mathrm{exp}(-S(Q,b))$.

Although Eq.~(\ref{eq:fs}) provides the general structure for the evolution of TMD distributions in $b$ space, it is only possible to calculate the $b$ dependence of $\mathcal{F}$ perturbatively in the small $b$ region.
In the large $b$ region, the $b$-dependence of $\mathcal{F}$ turns to be nonperturbative.
A convenient way to take into account the evolution behavior of $\tilde{F}(x,b;Q)$ in the large $b$ region is to include a nonperturbative Sudakov-like form factor $S_\mathrm{NP}$.
The latter one is usually given in a parameterized form, which can be obtained by fitting from the experimental data.
To allow a smooth transition of $b$ from perturbative region to nonperturbative region as well as to avoid the hitting on the Landau pole, one can set a parameter $b_{\mathrm{max}}$ to be the boundary between the two different regions.
The typical value of $b_{\mathrm{max}}$ is chosen around $1\ \mathrm{GeV}^{-1}$ to guarantee that $b_{\ast}$ is always in the perturbative region.
A $b$-dependent function $b_\ast(b)$ may be also introduced to have the property $b_\ast\approx b$ at small $b$ value and $b_{\ast}\approx b_{\mathrm{max}}$ at large $b$ value.
There are several different choices on the form of $b_\ast(b)$~\cite{Collins:1984kg,Collins:2016hqq,Bacchetta:2017gcc}.
In this work we choose it as $b_\ast=b/\sqrt{1+b^2/b_{\rm max}^2}  \ ,~b_{\rm max}<1/\Lambda_\mathrm{QCD}$~\cite{Collins:1984kg,Kang:2015msa}.

Combining the perturbative part and the nonperturbative part, one has the complete result for the Sudakov form factor appearing in Eq.~(\ref{eq:fs})
\begin{equation}
\label{eq:S}
S(Q,b)=S_{\mathrm{P}}(Q,b)+S_{\mathrm{NP}}(Q,b).
\end{equation}
with the boundary of the two parts set by the $b_\mathrm {max}$.
The perturbative part $S_{\mathrm{P}}(Q,b)$ has been studied~\cite{Echevarria:2014xaa,Kang:2011mr,Aybat:2011ge,Echevarria:2012pw,Echevarria:2014rua} in details and has the following form:
\begin{equation}
\label{eq:Spert}
S_{\mathrm{P}}(Q,b)=\int^{Q^2}_{\mu_b^2}\frac{d\bar{\mu}^2}{\bar{\mu}^2}\left[A(\alpha_s(\bar{\mu}))
\mathrm{ln}\frac{Q^2}{\bar{\mu}^2}+B(\alpha_s(\bar{\mu}))\right],
\end{equation}
which is the same for different kinds of distribution functions, namely, $S_{P}$ is spin-independent. In addition, the coefficients $A$ and $B$ in Eq.(\ref{eq:Spert}) can be expanded as the series of $\alpha_s/{\pi}$:
\begin{align}
A=\sum_{n=1}^{\infty}A^{(n)}(\frac{\alpha_s}{\pi})^n,\\
B=\sum_{n=1}^{\infty}B^{(n)}(\frac{\alpha_s}{\pi})^n.
\end{align}
In this work, we will take $A^{(n)}$ up to $A^{(2)}$ and $B^{(n)}$ up to $B^{(1)}$ in the accuracy of next-to-leading-logarithmic (NLL) order~\cite{Collins:1984kg,Landry:2002ix,Qiu:2000ga,Kang:2011mr,Aybat:2011zv,Echevarria:2012pw} :
\begin{align}
A^{(1)}&=C_F,\\
A^{(2)}&=\frac{C_F}{2}\left[C_A\left(\frac{67}{18}-\frac{\pi^2}{6}\right)-\frac{10}{9}T_Rn_f\right],\\
B^{(1)}&=-\frac{3}{2}C_F.
\end{align}

A general form of the nonperturbative part of the Sudakov form factor $S_{\rm NP}(Q;b)$ was suggested in Ref.~\cite{Collins:1984kg}:
\begin{equation}
\label{eq:snp_gene}
S_{\rm NP}(Q;b)=g_2(b)\ln Q/Q_0 +g_1(b).
\end{equation}
Here, $g_i(b)$ are the functions of the impact parameter $b$.
Particularly, $g_2(b)$ contains the information on the large $b$ behavior of the evolution kernel $\tilde K$, while $g_1(b)$ contains information about the intrinsic nonperturbative transverse motion of bound partons, i.e., it depends on the type of the hadron and quark flavor. It might also depend on the momentum fraction of the partons $x$~\cite{Su:2014wpa}.
It is also worth pointing out that $g_2(b)$ is universal for different types of TMDs and does not depend on the particular process, which is one of the important predictions of QCD factorization
theorems involving TMDs~\cite{Collins:2011zzd,Aybat:2011zv,Echevarria:2014xaa,Kang:2015msa}.

For $S_{\mathrm{NP}}$ associated with the $pp$ collision, a parametrization that can describe the SIDIS and Drell-Yan data with $Q$ values ranging from a few
to ten GeV has been proposed in Ref.~\cite{Su:2014wpa}
\begin{align}
S_{\mathrm{NP}}=g_1b^2+g_2\mathrm{ln}\frac{b}{b_{\ast}}\mathrm{ln}\frac{Q}{Q_0}
+g_3b^2\left((x_0/x_1)^{\lambda}+(x_0/x_2)^\lambda\right).
\label{eq:SNP_DY_NN}
\end{align}
The parameters $g_i$ are fitted from the nucleon-nucleon Drell-Yan process data~
\cite{Ito:1980ev, Antreasyan:1981uv,Moreno:1990sf,Affolder:1999jh,Abbott:1999wk,Abazov:2007ac,Aaltonen:2012fi}
at the initial scale of $Q^2_0=2.4\ \mathrm{GeV}^2$ yielding $g_1=0.212,\ g_2=0.84, \ g_3 = 0$.
Since the nonperturbative form factor $S_{\mathrm{NP}}$ for quarks from the one proton and antiquarks from another proton satisfies~\cite{Prokudin:2015ysa}
\begin{align}
S^q_{\mathrm{NP}}(Q,b)+S^{\bar{q}}_{\mathrm{NP}}(Q,b)=S_{\mathrm{NP}}(Q,b),
\end{align}
$S_{\mathrm{NP}}$ associated with a single TMD distribution function can be expressed as
\begin{align}
\label{eq:SNPproton}
S^{f_{1,q/p}}_{\mathrm{NP}}(Q,b)=\frac{g_1}{2}b^2+\frac{g_2}{2}\ln\frac{b}{b_{\ast}}\ln\frac{Q}{Q_0}.
\end{align}
In our calculation of the pion-proton Drell-Yan process, we will adopt the above form factor for the unpolarized TMD distributions of the proton.

For the nonperturbative form factors of the pion distribution function, we adopt the parametrization proposed in Ref.~\cite{Wang:2017zym}
\begin{align}
S^{f_{1,q/\pi}}_{\mathrm{NP}}=g^\pi_1\,b^2+g^\pi_2\mathrm{ln}\frac{b}{b_{\ast}}\mathrm{ln}\frac{Q}{Q_0},
\label{eq:SNP_pion}
\end{align}
which has the same form as that for the proton (in the case $g_3=0$).
After fitting to the $\pi^- N$ Drell-Yan data ~\cite{Conway:1989fs}, the values of the parameters $g_1^\pi$ and $g_2^\pi$ are obtained at the initial energy scale $Q^2_0=2.4\ \mathrm{GeV}^2$ as $g^\pi_1=0.082$ and $g^\pi_2=0.394$.
In the fit we also chose $b_{\mathrm{max}}=1.5\ \mathrm{GeV}^{-1}$, in consistence with the choice in Ref~.\cite{Su:2014wpa}.
We note that a form of $S^{f_{1,q/\pi}}_{\mathrm{NP}}$ motivated by the NJL model was given in Ref.~\cite{Ceccopieri:2018nop}.

Besides the Sudakov form factor in Eq.~(\ref{eq:fs}), another important element in Eq.~(\ref{eq:fs}) is the TMD distribution function at a fixed scale $\tilde{F}(x,b,\mu)$.
In the small $b$ region $1/Q \ll b \ll 1/ \Lambda$, the TMD distributions at a fixed scale $\mu$ can be expressed as the convolution of the perturbatively calculable hard coefficients $C$ and the corresponding collinear counterparts, which could be the collinear PDFs or the multiparton correlation functions~\cite{Collins:1981uk,Bacchetta:2013pqa}
\begin{equation}
\tilde{F}_{q/H}(x,b;\mu)=\sum_i C_{q\leftarrow i}\otimes F_{i/H}(x,\mu).
\label{eq:small_b_F}
\end{equation}
The convolution $\otimes$ regarding the momentum fraction of $x$ is given by
\begin{equation}
 C_{q\leftarrow i}\otimes F_{i/H}(x,\mu)\equiv \int_{x}^1\frac{d\xi}{\xi} C_{q\leftarrow i}(x/\xi,b;\mu)F_{i/H}(\xi,\mu),
 \label{eq:otimes}
\end{equation}
and $F_{i/H}(\xi,\mu)$ is the corresponding collinear counterpart of the TMD distribution of flavor $i$ in hadron $H$ at the energy scale $\mu$, which could be a dynamic scale related to $b_*$ by $\mu_b=c_0/b_*$, with $c_0=2e^{-\gamma_E}$ and the Euler Constant $\gamma_E\approx0.577$~\cite{Collins:1981uk}.
$\sum_i$ is the sum of both quark and antiquark flavors.

It is straightforward to rewrite the scale-dependent TMD distribution function $\tilde{F}$ of the proton and the pion in $b$ space
\begin{align}
\label{eq:tildeF}
\tilde{F}_{q/H}(x,b;Q)=e^{-\frac{1}{2}S_{\mathrm{P}}(Q,b_\ast)-S^{F_{q/H}}_{\mathrm{NP}}(Q,b)}
\mathcal{F}(\alpha_s(Q))\sum_i C^F_{q\leftarrow i}\otimes F_{i/H}(x,\mu_b),
\end{align}
The factor of $\frac{1}{2}$ in front of $S_{\mathrm{P}}$ comes from the fact that $S_{\mathrm{P}}$ of quarks and antiquarks satisfies the relation ~\cite{Prokudin:2015ysa}
\begin{align}
S^q_{\mathrm{P}}(Q,b_\ast)=S^{\bar{q}}_{\mathrm{P}}(Q,b_\ast)=S_{\mathrm{P}}(Q,b_\ast)/2.
\end{align}

The hard coefficients $C^F_{q\leftarrow i}$ and $\mathcal{F}$ for $f_1$ and $h_1$ have been calculated up to next-to-leading order (NLO), while those for the Boer-Mulders function still remain in the leading order (LO).
For consistency, in this work we will adopt the LO results of the $C$ coefficients for  $f_{1}$, $h_{1}^{\perp}$ and $h_{1}$. That is, we take $\mathcal{F}=1$ and $C^F_{q\leftarrow i}=\delta_{qi}\delta(1-x)$ for $F=f_1, h_1$ and $h_1^{\perp}$. We also note that a calculation in Ref.~\cite{Kang:2015msa} shows that the NLO $C$-coefficient for $h_1$ vanishes.

With all the ingredients above, we can obtain the unpolarized distribution function of the proton and pion in $b$ space as
\begin{align}
\tilde{f}_{1,q/p}(x,b;Q) &=e^{-\frac{1}{2}S_{\mathrm{P}}(Q,b_\ast)-S^{f_{1, q/p}}_{\mathrm{NP}}(Q,b)}
 f_{1, q/p}(x,\mu_b),\nonumber\\
\tilde{f}_{1,q/\pi}(x,b;Q) &=e^{-\frac{1}{2}S_{\mathrm{P}}(Q,b_\ast)-S^{f_{1, q/\pi}}_{\mathrm{NP}}(Q,b)}
 f_{1, q/\pi}(x,\mu_b).
\label{eq:f_b}
\end{align}

The distribution function in the transverse momentum space can be obtained by performing the Fourier transformation on the $\tilde{f}_{1,q/H}(x,b;Q)$
\begin{align}
f_{1,q/p}(x,k_\perp;Q)=\int_0^\infty\frac{db b}{2\pi}J_0(k_\perp b)e^{-\frac{1}{2}S_{\mathrm{P}}(Q,b_\ast)-S^{f_{1, q/p}}_{\mathrm{NP}}(Q,b)} f_{1, q/p}(x,\mu_b),\label{eq:f_proton}\\
f_{1,q/\pi}(x,k_\perp;Q)=\int_0^\infty\frac{db b}{2\pi}J_0(k_\perp b)e^{-\frac{1}{2}S_{\mathrm{P}}(Q,b_\ast)-S^{f_{1, q/\pi}}_{\mathrm{NP}}(Q,b)} f_{1, q/\pi}(x,\mu_b),
\end{align}
where $J_0$ is the Bessel function of the first kind, and $k_\perp = |\bm k_\perp|$.

Similar to the unpolarized distribution function, the transversity distribution of the proton in $b$-space and $k_\perp$ space can be obtained as~\cite{Kang:2015msa}
\begin{align}
&\widetilde{h}_{1,q/p}(x,b;Q)=e^{-\frac{1}{2}S_{\mathrm{P}}(Q,b_\ast)-S^{f_{1,q/p}}_{\mathrm{NP}}(Q,b)}
h_{1,q/p}(x,\mu_b),\label{eq:h_proton_b}\\
&h_{1,q/p}(x,k_\perp;Q)=\int_0^\infty\frac{dbb}{2\pi}J_0(k_\perp b)e^{-\frac{1}{2}S_{\mathrm{P}}(Q,b_\ast)-S^{f_{1,q/p}}_{\mathrm{NP}}(Q,b)}
h_{1,q/p}(x,\mu_b), \label{eq:h_p_k}
\end{align}
where the factors and coefficients related to the hard scattering are adopted at LO and the corresponding collinear distribution is the integrated transversity $h_1(x)$. The nonperturbative Sudakov form factor associated with the proton transversity distribution is also assumed to be the same as that for unpolarized distribution function~\cite{Kang:2015msa}.

According to Eq.~(\ref{eq:small_b_F}), in the small $b$ region, we can also express the Boer-Mulders function of the pion beam at a fixed energy scale $\mu$ in terms of the perturbatively calculable coefficients and the corresponding collinear correlation function
\begin{align}
\widetilde{h}_{1,q/\pi}^{\alpha\perp}(x,b;\mu)=(\frac{-ib_\perp^\alpha}{2})T^{(\sigma)}_{q/\pi,F}(x,x;\mu),
\end{align}
where the hard coefficients are calculated up to LO, and the Boer-Mulders function in the $b$ space is defined as
\begin{align}
\tilde{h}_{1,q/\pi}^{\perp\alpha(\mathrm{DY})}(x,b;\mu)=\int d^2\bm{k}_\perp e^{-i\bm{k}_\perp\cdot\bm{b}_\perp}\frac{k^\alpha_\perp}{M_\pi}
h^{\perp(\mathrm{DY})}_{1,q/\pi}(x,\bm{k}^2_\perp;\mu). \label{eq:pibm}
\end{align}
The collinear function $T^{(\sigma)}_{q/\pi,F}(x,x;\mu)$ is the chiral-odd twist-3 quark-gluon-quark correlation function, which is related to the first transverse moment of the Boer-Mulders function $h_{1,q/\pi}^{\perp (1)}$ by
\begin{align}
T^{(\sigma)}_{q/\pi,F}(x,x;\mu)=\int d^2 \bm{k}_\perp\frac{\bm{k}_\perp^2}{M_\pi}h_{1,q/\pi}^\perp(x,\bm{k}^2_\perp;\mu)
= 2M_\pi h_{1,q/\pi}^{\perp (1)}. \label{eq:qsbm}
\end{align}

As for the nonperturbative part of the Sudakov form factor associated with the Boer-Mulders function,
the information still remains unknown. In a practical calculation, we assume that it is the
same as $S_\mathrm{NP}^{f_{1,q/\pi}}$, i.e., $S_\mathrm{NP}^{h_{1,q/\pi}^\perp}=S_\mathrm{NP}^{f_{1,q/\pi}}$.
Therefore, we can obtain the Boer-Mulders function of the pion in $b$-space as
\begin{align}
\widetilde{h}_{1,q/\pi}^{\alpha\perp}(x,b;Q)=(\frac{-ib_\perp^\alpha}{2})
e^{-\frac{1}{2}S_{\mathrm{P}}(Q,b_\ast)-S^{f_{1,q/\pi}}_{\mathrm{NP}}(Q,b)}
T^{(\sigma)}_{q/\pi,F}(x,x;\mu_b).
\label{eq:BM_b}
\end{align}

After performing the Fourier transformation back to the transverse momentum space, one can get the Boer-Mulders function as
\begin{align}
\frac{k_\perp}{M_\pi}h^\perp_{1,q/\pi}(x,k_\perp;Q)=\int_0^\infty db(\frac{b^2}{2\pi})J_1(k_\perp b)e^{-\frac{1}{2}S_{\mathrm{P}}(Q,b_\ast)-S^{f_{1, q/\pi}}_{\mathrm{NP}}(Q,b)}
h^{\perp(1)}_{1,q/\pi}(x;\mu_b).
\label{eq:BM_kt}
\end{align}

\section{Formalism of the ${\sin (2\phi-\phi_S)}$ asymmetry in Drell-Yan process}

\label{Sec.formalism}

In this section, we present the formalism of the ${\sin (2\phi-\phi_S)}$ asymmetry in Drell-Yan process within TMD factorization following the procedure in Ref.~\cite{Collins:2011zzd}. We take into account the TMD evolution effects to obtain the theoretical expression of the ${\sin (2\phi-\phi_S)}$ asymmetry, which arises from the convolution of Boer-Mulders function of the pion beam and the transversity distribution function of the proton target at the leading twist.

The process we study is the pion-induced Drell-Yan process
\begin{align}
\pi^-(P_\pi)+p^\uparrow(P_p)\longrightarrow \gamma^*(q)+X \longrightarrow l^+(\ell)+l^-(\ell')+X,
\end{align}
where $P_\pi$, $P_p$ and $q$ stand for the four-momenta of the incoming $\pi^-$ meson, the proton target and the virtual photon, respectively, $Q^2=q^2$ is the invariant mass square of the lepton pair, and $\uparrow$ denotes the transverse polarization of the target.
We adopt the following kinematical variables to express the experimental observables~\cite{Collins:1984kg,Gautheron:2010wva}
\begin{align}
&s=(P_{\pi}+P_p)^2,\quad x_\pi=\frac{Q^2}{2P_\pi\cdot q},\quad x_p=\frac{Q^2}{2P_p\cdot q},\nonumber\\
&x_F=2q_L/s=x_\pi-x_p,\quad\tau=Q^2/s=x_\pi x_p,\quad y=\frac{1}{2}\mathrm{ln}\frac{q^+}{q^-}=\frac{1}{2}\mathrm{ln}\frac{x_\pi}{x_p},
\end{align}
where $s$ is the total center-of-mass~(c.m.) energy squared; $x_\pi$ and $x_p$ are the Bjorken variables of the pion and proton, respectively; $q_L$ is the longitudinal momentum of the virtual photon in the c.m. frame of the incident hadrons; $x_F$ is the Feynman $x$ variable; and $y$ is the rapidity of the lepton pair. Thus, $x_\pi$ and $x_p$ can be expressed as functions of $x_F$, $\tau$ and of $y$, $\tau$
\begin{align}
x_{\pi/p}=\frac{\pm x_F+\sqrt{x_F^2+4\tau}}{2},\quad x_{\pi/p}=\sqrt{\tau} e^{\pm y}.
\end{align}

In leading twist, the differential cross section in $\pi p$ Drell-Yan for a transversely polarized target has the following general form~\cite{Gautheron:2010wva}
\begin{align}
&\frac{d\sigma}{d^4qd\Omega}=\frac{\alpha_{em}^2}{F q^2}\hat{\sigma}_U \Big\{\Big(1+D_{[\sin^2\theta]}A_U^{\cos2\phi}\cos2\phi\Big)\nonumber\\
&\quad\quad\quad\quad\quad\quad\quad\quad+ |\bm{S}_T|\Big[A_T^{\sin\phi_S}\sin\phi_S+D_{[\sin^2\theta]}
\Big(A_T^{\sin(2\phi+\phi_S)}\sin(2\phi+\phi_S)\nonumber\\
&\quad\quad\quad\quad\quad\quad\quad\quad
+A_T^{\sin(2\phi-\phi_S)}\sin(2\phi-\phi_S)
\Big)\Big]\Big\}.
\end{align}
Here, $\phi_S$ represents the azimuthal angle of the target polarisation vector $S_T$ in the target rest frame, $\phi$ and $\theta$ denote the azimuthal and polar angles of the lepton momentum in the Collins-Soper frame, $\hat{\sigma}_U = F_U^1(1+\cos^2\theta)$, with $F_U^1$ the unpolarized structure function.
The symbol $D_{[f(\theta)]}$ denotes the depolarization factor that depends on $\theta$ only, and in LO it is simplified to $\sin^2\theta/(1+\cos^2\theta)$.
Furthermore, $A_{P}^{f[\phi,\phi_S]}$ denotes the azimuthal asymmetry with a modulation of $f[\phi,\phi_S]$, where $P=U$ or $T$ denotes the polarization of the target proton~($U$ for unpolarized, while $T$ for transversely polarized).
The asymmetry $A_{P}^{f[\phi,\phi_S]}$ can be written as the ratio between the corresponding structure function $F_{P}^{f[\phi,\phi_S]}$ and the unpolarized structure function.
In this work, we focus on the ${\sin (2\phi-\phi_S)}$ asymmetry:
\begin{align}
A_{T}^{\sin (2\phi-\phi_S)}(x_1,x_2,Q)=\frac{F_{T}^{\sin (2\phi-\phi_S)}(x_1,x_2,Q)}{F_{U}^{1}(x_1,x_2,Q)}.
\label{eq:asy}
\end{align}
The denominator can expressed as the convolution of the unpolarized distribution functions from each hadron
\begin{align}
F_{U}^{1}&=\mathcal{C}[f_{1,q/\pi} f_{1,\bar{q}/p}],
\label{eq:FUU}
\end{align}
while the numerator ($\bm{h}=\hat{\bm q}\equiv \bm{q}_\perp/{|\bm{q}_\perp|}$)~\cite{Boer:1999mm,Arnold:2008kf}
\begin{align}
F_{T}^{\sin (2\phi-\phi_S)}&=-\mathcal{C}[\frac{\bm{h}\cdot\bm{k}_{a\perp}}{M_\pi}h_{1,q/\pi}^\perp h_{1,\bar{q}/p}]
\label{eq:FUT}
\end{align}
is the convolution of the pion Boer-Mulders distribution and the proton transversity distribution.
The convolution of TMDs in the transverse momentum space is defined through the following notation
\begin{align}
\mathcal{C}[\omega(\bm{k}_{a\perp},\bm{k}_{b\perp})f_{1}\bar{f}_{2}]
=\frac{1}{N_{c}}\sum_{q}e_{q}^{2}\int d^{2}\bm{k}_{a\perp}d^{2}\bm{k}_{b\perp}\delta^{2}(\bm{k}_{a\perp}+\bm{k}_{b\perp}-\bm{q}_\perp)
\omega(\bm{k}_{a\perp},\bm{k}_{b\perp})\times    \nonumber\\
  \Big[f^{q}_{1}(x_a,\bm{k}^2_{a\perp})f^{\bar{q}}_{2}(x_b,\bm{k}^2_{b\perp})
+f^{\bar{q}}_{1}(x_a,\bm{k}^2_{a\perp})f^{q}_{2}(x_b,\bm{k}^2_{b\perp})\Big],
\label{eq:C}
\end{align}
with $N_c=3$ being the the number of colors, $\bm{q}_\perp, \bm{k}_{a\perp}$, and $\bm{k}_{b\perp}$ denoting the transverse momenta of the lepton pair, quark and antiquark in the initial hadrons. Finally, $\omega(\bm{k}_{a\perp},\bm{k}_{b\perp})$ is an arbitrary function of $\bm{k}_{a\perp}$ and $\bm{k}_{b\perp}$.

In general, it is more convenient to study the structure function first in the $b$-space, in which the convolution of the TMD distributions can be resolved to the product of $b$-dependent TMDs.
The physical observables can be obtained through a Fourier transformation from the $b$-space to the $k_\perp$-space.
Using the property of the following Fourier transformation
\begin{align}
\delta^{2}(\bm{k}_{a\perp}+\bm{k}_{b\perp}-\bm{q}_\perp) = {1\over (2\pi)^2}\int d^2 \bm{b}_\perp e^{- i \bm b_\perp\cdot(
\bm{k}_{a\perp}+\bm{k}_{b\perp}-\bm{q}_\perp)},
\end{align}
One can obtain the spin-dependent structure function $F_{T}^{\sin(2\phi-\phi_S)}$ as
\begin{align}
F_{T}^{\sin(2\phi-\phi_S)}
=& -\frac{1}{N_c}\sum_q e_q^2 \int d^{2}\bm{k}_{a\perp}d^{2}\bm{k}_{b\perp}\int\frac{d^2\bm{b}_\perp}{(2\pi)^2}
e^{-i\bm{b}_\perp\cdot(\bm{k}_{a\perp}+\bm{k}_{b\perp}-\bm{q}_\perp)}
\frac{\bm{h}\cdot\bm{k}_{a\perp}}{M_\pi}h_{1,q/\pi}^\perp(x_\pi,\bm{k}^2_{a\perp})
h_{1,\bar{q}/p}(x_p,\bm{k}^2_{b\perp})
+(q\leftrightarrow \bar{q})
\nonumber\\
=&-\frac{1}{N_c}\sum_{q} e_q^2\int_0^\infty \frac{db}{{4\pi}}b^2J_1(q_\perp b)
h_{1,q/p}(x_p,\mu_b)
T^{(\sigma)}_{\bar{q}/\pi,F}(x_\pi,x_\pi,\mu_b)
e^{-\left(S^{f_{1,q/p}}_{\mathrm{NP}}+S^{f_{1,q/\pi}}_{\mathrm{NP}}+S_\mathrm{P}\right)}
+ (q\leftrightarrow \bar{q}),
\label{eq:final FT}
\end{align}
where, we have used Eqs.~(\ref{eq:h_p_k}), (\ref{eq:pibm}) and (\ref{eq:qsbm}).
The unpolarized structure function can be expressed in a similar way:
\begin{align}
F_{U}^{1}&=\frac{1}{N_c}\sum_q e_q^2 \int d^{2}\bm{k}_{a\perp}d^{2}\bm{k}_{b\perp}\int\frac{d^2\bm b_\perp}{(2\pi)^2}
e^{-i(\bm{k}_{a\perp}+\bm{k}_{b\perp}-\bm{q}_\perp)\cdot\bm{b_\perp}}
f_{1,q/\pi}(x_\pi,\bm{k}^{2}_{a\perp})f_{1,\bar{q}/p}(x_p,\bm{k}^{2}_{b\perp})
\nonumber\\
&=\frac{1}{N_c}\sum_q e_q^2\int_0^\infty \frac{b db}{2\pi}J_0(q_\perp b)f_{1,q/\pi}(x_\pi,\mu_b)f_{1,\bar{q}/p}(x_p,\mu_b)
e^{-\left(S^{f_{1,q/p}}_{\mathrm{NP}}+S^{f_{1,q/\pi}}_{\mathrm{NP}}+S_\mathrm{P}\right)}
+ (q\leftrightarrow \bar{q}),
\label{eq:final_FU}
\end{align}
where the expression of the unpolarized distribution function in Eq.~(\ref{eq:f_b}) is included and the definition of the unpolarized distribution function in $b$-space is
\begin{equation}
\tilde{f}_{1,q/H}(x_H,b;\mu)=\int d^2\bm{k}_\perp e^{-i \bm{b}_\perp\cdot\bm{k_\perp}}f_{1,q/H}(x_H,\bm{k^2_\perp};\mu).
\end{equation}

\section{Numerical calculation}

\label{Sec.numerical}

Using the framework set up above, in this section we present the numerical calculation of the $\sin (2\phi-\phi_S)$ azimuthal asymmetry in the pion-induced transversely polarized Drell-Yan process. We estimate the asymmetry at the kinematics of the COMPASS Drell-Yan program and compare it with the recent COMPASS measurement.
To do this we need to know the corresponding distribution functions of the pion meson, as well as those of the proton target.
For the former one, as there is no extraction on the Boer-Mulders function of the pion meson, we apply a model result based on the light-cone wave function of the pion meson from Ref.~\cite{Wang:2017onm} at the model scale $\mu_0^2=0.25\mathrm{GeV}^2$:
\begin{align}
{h}^{\perp }_{1,\pi }(x,\bm{k}^{2}_\perp)&=\frac{{C}_{F}{\alpha }_{s}}{16{\pi }^{3}}\frac{mM_\pi}{\sqrt{{m}^{2}+\bm{k}^{2}_\perp}}\frac{{A}^{2}}{\bm{k}^{2}_\perp}\mathrm{exp}
[-\frac{1}{8{\beta }^{2}}\frac{\bm{k}^{2}_\perp+{m}^{2}}{x(1-x)}]\left[\Gamma (\frac{1}{2},\frac{{m}^{2}}{8\beta^2 x(1-x)})-\Gamma (\frac{1}{2},\frac{\bm{k}^{2}_\perp+{m}^{2}}{8\beta^2 x(1-x)})\right],
\end{align}
where the values of the parameters are~\cite{Xiao:2003wf,Wang:2017onm},
\begin{align}
\beta=0.41\ \textrm{GeV},\quad m_u=m_d=m=0.2\ \textrm{GeV}, \quad A=31.303\ \textrm{GeV}^{-1}.
\end{align}
The corresponding collinear twist-3 distribution $T^{\sigma}_{q,F}(x,x,\mu_0)$ at the model scale can be obtained by using Eq.~(\ref{eq:qsbm}).
For consistency, we apply the unpolarized distribution function of the pion meson $f_{1\pi}(x)$ using the same model.

\begin{figure}
  \centering
  \includegraphics[width=0.48\columnwidth]{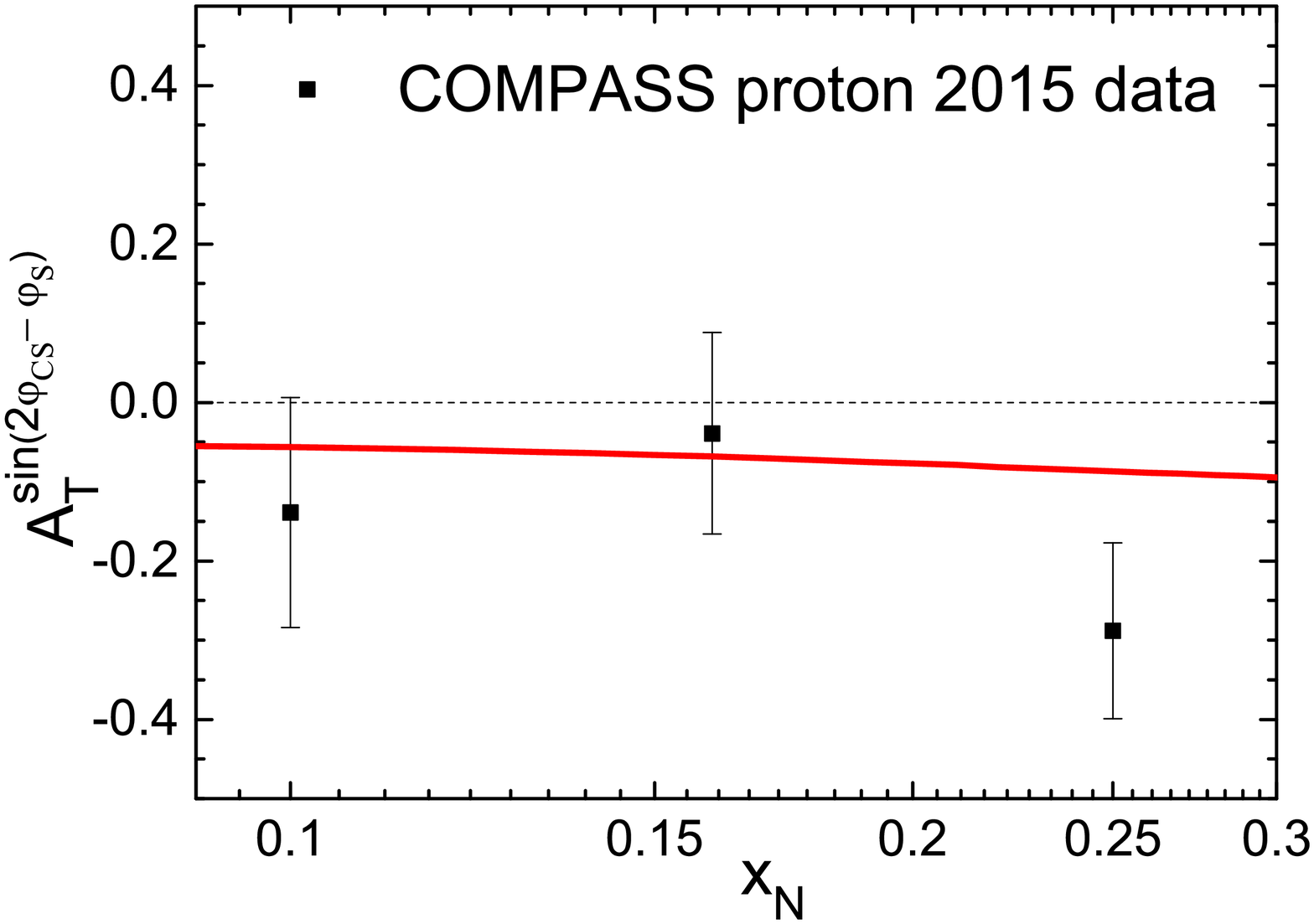}
  \includegraphics[width=0.48\columnwidth]{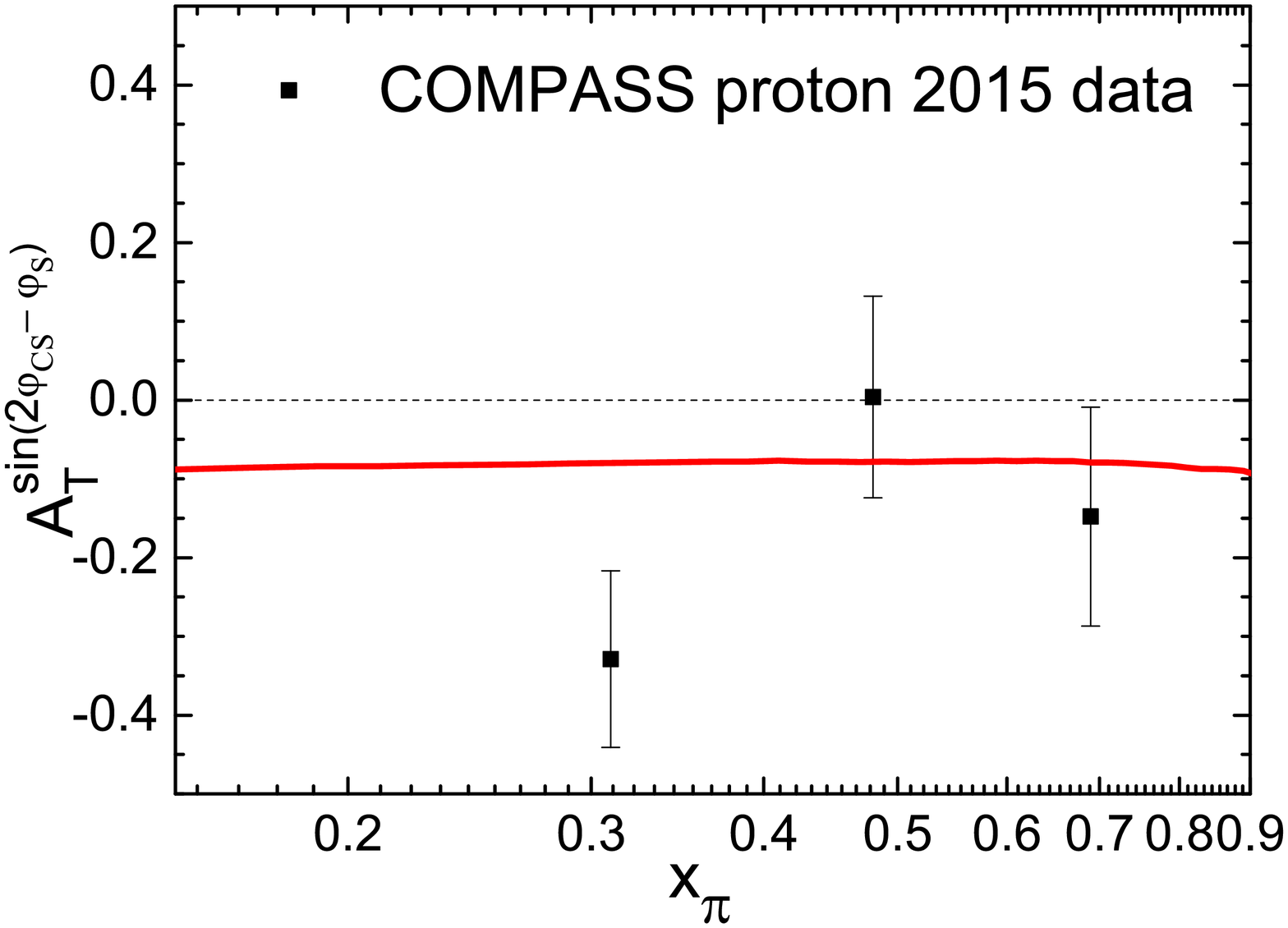}\\
  \includegraphics[width=0.48\columnwidth]{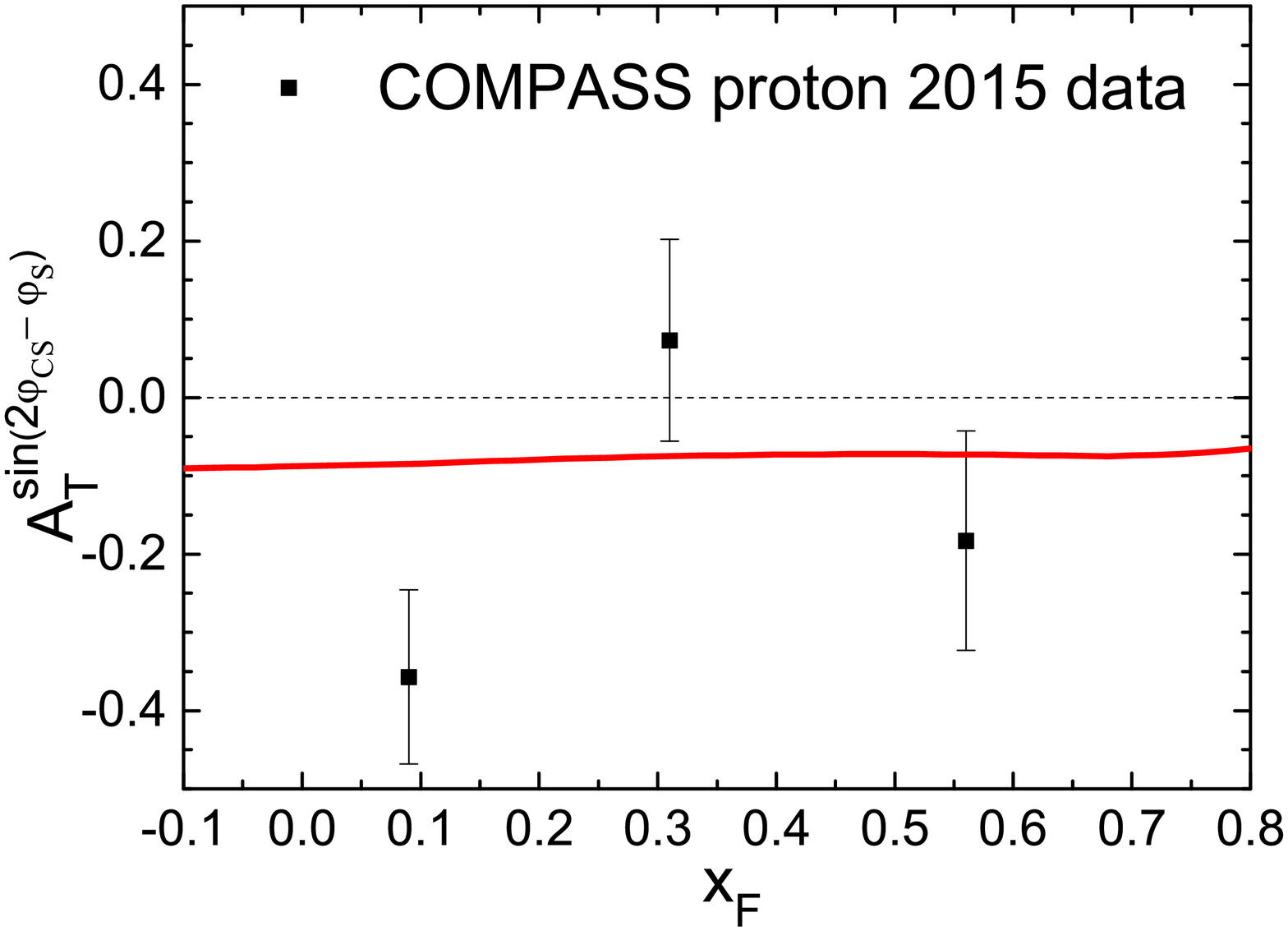}
  \includegraphics[width=0.48\columnwidth]{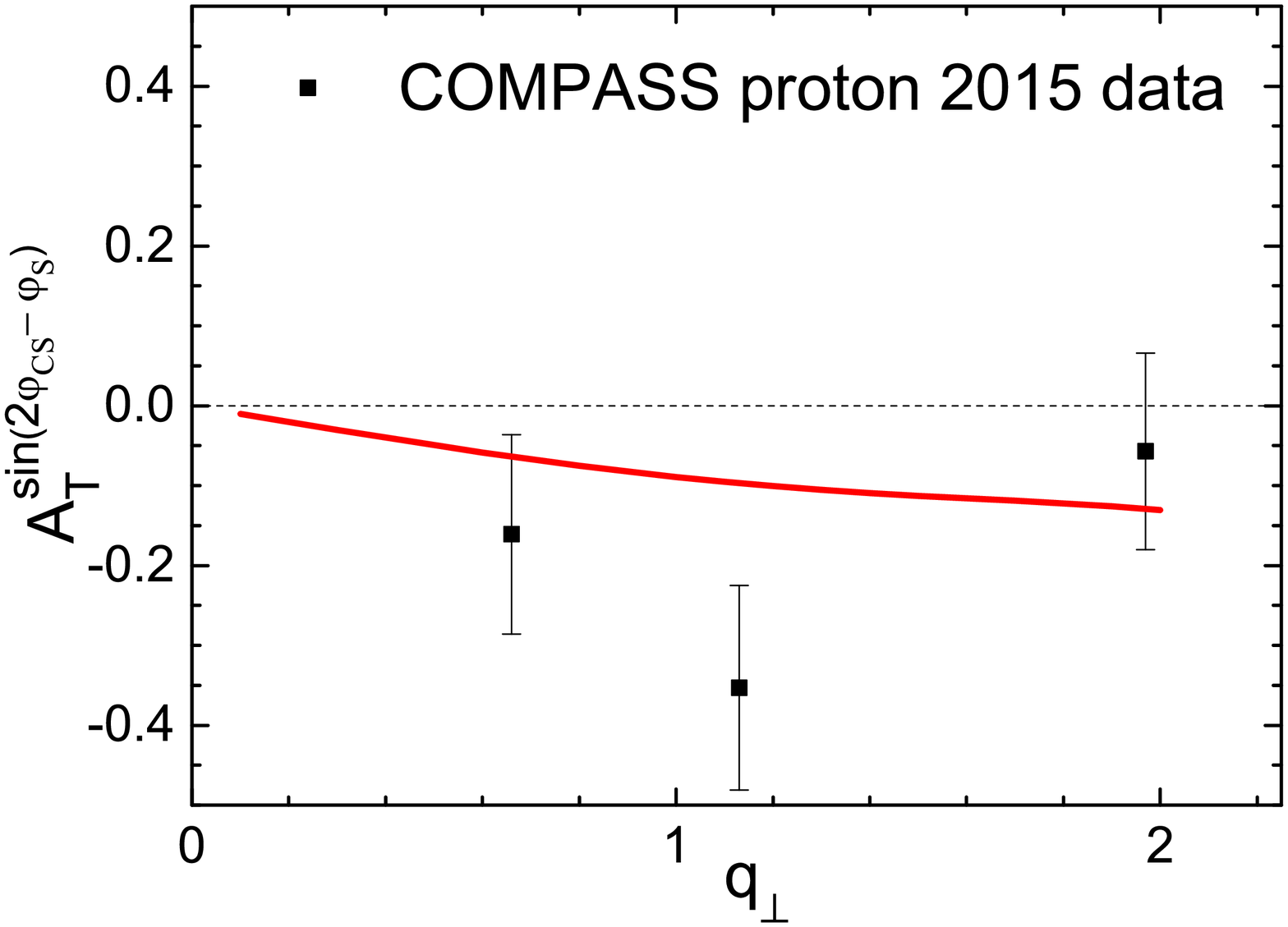}
  \caption{The $\sin (2\phi-\phi_S)$ azimuthal asymmetry for $\pi^-$ scattering off transversely polarized Drell-Yan process as functions of $q_\perp$(upper left), $x_F$(upper right), $x_N$(lower left) and $x_\pi$(lower right), compared with the COMPASS data.}
  \label{fig:asy}
\end{figure}

For the collinear distributions of the proton, we resort to existing parametrizations, i.e., we adopt the NLO set of the CT10 parametrization~\cite{Lai:2010vv}~(central PDF set) for the unpolarized distribution function $f_1(x)$ of the proton, and we choose the transversity distribution extracted from SIDIS data~\cite{Kang:2015msa} via the same TMD evolution formalism:
\begin{align}
h_1^q(x,Q_0)=N_q^hx^{a_q}(1-x)^{b_q}\frac{(a_q+b_q)^{a_q+b_q}}{a_q^{a_q}b_q^{b_q}}\times \frac{1}{2}(f_1^q(x,Q_0)+g_1^q(x,Q_0)),
\label{eq:trans-para}
\end{align}
where $g_1^q$ is helicity distribution function~\cite{deFlorian:2009vb}.

We apply the {\sc{QCDNUM}} evolution package~\cite{Botje:2010ay} to perform the evolution of $f_{1,q/\pi}$ from the model scale $\mu_0$ to another energy.
As for the energy evolution of the twist-3 collinear correlation function $T^{(\sigma)}_{q,F}$, the evolution effect has been studied in Refs.~\cite{Kang:2012em,Kang:2008ey,Vogelsang:2009pj,Zhou:2008mz,Braun:2009mi}.
For simplicity, we only consider the homogenous term in the evolution kernel
\begin{align}
P^{T^{(\sigma)}_{q,F}}_{qq}(x)\approx\Delta_T\,P_{qq}(x)-N_C\delta(1-x),\label{eq:evobm}\\
\Delta_T\,P_{qq}(x)=C_F\left[\frac{2z}{(1-z)_+}+\frac{3}{2}\delta(1-x)\right],\label{eq:evh1}
\end{align}
with $\Delta_T\,P_{qq}$ being the evolution kernel for the transversity distribution function $h_1(x)$. We customize the original code of {\sc{QCDNUM}} to include the approximate kernel in Eq.~(\ref{eq:evobm}).
Similarly, we also include the kernel in Eq.~(\ref{eq:evh1}) to solve the DGLAP evolution equations for the transversity distribution function of proton.

The COMPASS Collaboration at CERN has reported the first measurement of the transverse-spin-dependent azimuthal asymmetries in the Drell-Yan process~\cite{Aghasyan:2017jop} in which a $\pi^-$ beam with $P_\pi=\ 190\ \mathrm{GeV}$ collides on a polarized $\mathrm{NH}_3$ target~\cite{Gautheron:2010wva,Aghasyan:2017jop}~(which can serve as a transversely polarized nucleon target). The covered kinematical ranges are as follows
\begin{align}
&0.05<x_N<0.4,\quad 0.05<x_\pi<0.9, \quad
4.3\ \mathrm{GeV}<Q<8.5\ \mathrm{GeV},\quad s=357~\mathrm{GeV}^2,\quad -0.3<x_F<1.
\label{eq:cuts}
\end{align}

In Fig.~\ref{fig:asy}, we plot our numerical result of the $\sin (2\phi-\phi_S)$ azimuthal asymmetry as functions of  $x_N, x_\pi, x_F$ and $q_\perp$ in the pion-induced Drell-Yan process based on the TMD factorization formalism described in Eqs.~(\ref{eq:asy}), (\ref{eq:final_FU}), and (\ref{eq:final FT}) at the kinematics of COMPASS. To make the TMD factorization valid in the kinematic region $q_\perp\ll Q$, the integration over the transverse momentum $q_\perp$ is performed in the region of $0<q_\perp<2~\mathrm{GeV}$, which is the same as the cut in Ref.~\cite{Sun:2013hua}.
The upper panels of Fig.~\ref{fig:asy} show the asymmetries as functions of $x_p$ (left panel) and $x_\pi$ (right panel); and the lower panels depict the $x_F$-dependent and $q_\perp$-dependent asymmetries, respectively. In the figure we also show the experimental data measured by the COMPASS Collaboration~\cite{Aghasyan:2017jop} for comparison.

As shown in Fig.~\ref{fig:asy}, in all the cases the $\sin (2\phi-\phi_S)$ azimuthal asymmetry in the $\pi^- p$ Drell-Yan from our calculation is negative, in agreement with most of the data from COMPASS.
Our estimate also shows that the asymmetry changes slightly with the change of $x_N$, $x_\pi$ and $x_F$, and the magnitude of the $x_N$-, $x_\pi$- and $x_F$-dependent asymmetries is around 0.05 to 0.10.
For the $q_\perp$ asymmetry, we find that its magnitude is about 0.05 to 0.15 and moderately increases with increasing $q_\perp$ in the region $q_T<2 \mathrm{GeV}$. Our numerical estimates show that the $A_{UT}^{\sin (2\phi-\phi_S)}$ is sizable at the the kinematics of COMPASS and is qualitatively consistent with the COMPASS measurement after considered the uncertainties of the data.
Our study demonstrate that, with the current knowledge on the distributions of the proton, it is promising to apply the evolution formalism of TMD distributions to study the SSA contributed by the chiral-odd distributions at the kinematics of COMPASS.
Our calculation also indicates that the proton transversity distribution may be served as a probe to access the pion Boer-Mulders function as well as the corresponding nonperturbative Sudakov form factor in the context of the current formalism on the transversely polarized $\pi^- p$ Drell-Yan process.

\section{Conclusion}
\label{Sec.conclusion}

In this work, we applied the TMD factorization to study the $\sin (2\phi-\phi_S)$ azimuthal asymmetry in the single transversely polarized $\pi^- p$ Drell-Yan process that is accessible at COMPASS.
The asymmetry arises from the coupling of the Boer-Mulders function of the pion beam and the transversity distribution of the proton target.
We took into account the TMD evolution of the asymmetry by including the Sudakov form factor for the TMD distributions of the pion and proton.
The hard coefficients associated with the corresponding collinear functions are kept in the leading-order accuracy.
For the transversity distribution of the proton used in the study, we employed a recent parametrization for which the same TMD evolution effect is considered.
For the distributions of the pion meson, we chose the result from a model calculation incorporating the light-cone wave function approach.
As the nonperturbative Sudakov form factor associated with the pion Boer-Mulders function is still unknown,  we assume that it is the same as that of the unpolarized distribution function. The latter one has been extracted from the unpolarized $\pi N$ Drell-Yan data.
We then calculated the $\sin (2\phi-\phi_S)$ azimuthal asymmetry in the $\pi^- p$ Drell-Yan process at the kinematics of COMPASS.
Our analysis demonstrated that, within the framework of TMD evolution, the $\sin (2\phi-\phi_S)$ asymmetry at COMPASS can be qualitatively described (sign and magnitude) by the current analysis on the TMD distributions of the pion and the proton.
Furthermore, our study may provide a framework to access the Boer-Mulders function of the pion and the corresponding nonperturbative Sudakov form factor through transversely polarized $\pi p$ data.

\section*{Acknowledgements}
This work is partially supported by the NSFC (China) grants 11575043, 11847217 and 11120101004. X. Wang is supported by the China Postdoctoral Science Foundation under Grant No.~2018M640680.

\end{document}